# On the Interactions of $Ti_2AlC$, $Ti_3AlC_2$, $Ti_3SiC_2$ and $Cr_2AlC$ with Palladium at 900°C

G. W. Bentzel[1], M. Sokol[1], J. Griggs, A. C. Lang and M. W. Barsoum[*]

*Department of Materials Science & Engineering, Drexel University, Philadelphia, PA 19104, USA*

This work was supported by the Department of Energy's Office of Nuclear Energy University Program under Grant CFP-11-3231.

[*]Corresponding author: Email – barsoumw@drexel.edu

[1]These authors contributed equally to this work.

**Abstract:**

Herein we report on the reactivity between palladium, Pd, and the MAX phases, $Ti_2AlC$, $Ti_3AlC_2$, $Ti_3SiC_2$ and $Cr_2AlC$. Diffusion couples of Pd/MAX were heated to 900°C under uniaxial stress of ~ 20 MPa for 2, 4, and 10 h in a vacuum (< 1 Pa) hot press. The diffusion couples were examined using X-ray diffraction, scanning electron microscopy and energy-dispersive X-ray spectroscopy. After heating to 900°C for 10 h, the diffusion layer thicknesses in the $Ti_2AlC$/Pd, $Cr_2AlC$/Pd, $Ti_3AlC_2$/Pd and $Ti_3SiC_2$/Pd couples were found to be 35, 45, 105 and 410 µm, respectively. Thus, $Ti_2AlC$ is the most resistant to reaction, while $Ti_3SiC_2$ is least resistant, with $Cr_2AlC$ and $Ti_3AlC_2$, in between. In all cases, the reaction occurred by the diffusion of the A-group element into Pd, concomitant with Pd diffusion into the MAX phase. No diffusion of M and X atoms was detected.

**Introduction:**

The $M_{n+1}AX_n$ or MAX, phases are layered, hexagonal, early transition-metal carbides and nitrides, where n = 1, 2, or 3, "*M*" is an early transition metal, "*A*" is an A-group (mostly groups 13 and 14) element, and "*X*" is C and/or N.[1] The MAX phases can be further categorized by their *n* value as "211s" for $M_2AX$ (*n* = 1), "312s" for $M_3AX_2$ (*n* = 2) and "413s" for $M_4AX_3$ (*n* = 3) etc.[2]

The MAX phases combine some of the more attractive properties of ceramics and metals together, that can be traced back to their chemical composition and structure. As with most metals, they have and high thermal and electric conductivities,[3-5] are easily machinable,[6] and have excellent thermal shock resistance and damage tolerance.[3,5,7] Some of the most studied, such as $Ti_2AlC$ and $Ti_3SiC_2$ are lightweight, stiff and creep,[8-12] oxidation,[13-16] and fatigue[17] resistant. Due to those unique combination of properties, the MAX phases have been recently proposed for use in the nuclear industry.[18-22] However, the published results on the chemical stability of the MAX phases vis-à-vis materials for nuclear systems is limited. The corrosion resistant in different coolant environment was studied for certain MAX phases. For example, it has been shown that $Ti_3SiC_2$ has superb corrosion resistance to molten Pb and Pb-Bi alloys.[23-25] It was established that interaction between the MAX phases and fuel cladding material, such as Zircaloy-4, SiC, and pyrolytic graphite, results in A-group element diffusion into the cladding material.[26,27] Recently we showed that polycrystalline samples of $Ti_3SiC_2$, $Ti_2AlC$, $Cr_2AlC$ apparently did not react after exposure to molten Na heated to 750°C for a week in stainless steel sealed tubes, although the $Ti_3AlC_2$ sample did react.[28] Moreover, results from the first ever reported study of neutron irradiated bulk MAX phases,[29-33] have shown that $Ti_2AlC$ and $Ti_3SiC_2$ are promising for high temperature nuclear applications.

An aspect for which there has been no work to date is related to the interaction between fission products, such as tritium and palladium with diffusion barrier layers of the TRISO fuel design.[34] Currently SiC and pyrolytic C are used as diffusion barrier layers. However, when Pd evolves from the fuel, it becomes a main factor in SiC corrosion, leading to cladding failure.[35-38] For example, it has been reported that Pd nodule attack leads to local thinning of the SiC layer.[37,38] If the Pd penetrates through the SiC, that would create a path for other fission products to escape, which is unacceptable.[37]

Other binary carbides and nitrides have also been tested to compare their resistance to Pd vs. SiC.[39,40] Tan et al.[39] found that TiN and ZrN, followed by TiC and ZrC, displayed better resistance to Pd attack than SiC. After heating to 1400°C for 10 h in a high purity Ar atmosphere, the formation of $TiPd_{(3+x)}$, $ZrPd_3$, and C were found in TiC/Pd and ZrC/Pd diffusion couples, in their respective systems. No intermediate phases were observed in a TiN/Pd diffusion couple, but a small amount of $ZrPd_3$ and C was observed in a ZrN/Pd system.[39] Demkowicz et al.[40] also tested TiN, TiC, and SiC with Pd, up to temperatures of 1600°C, and, again, found that both TiC and TiN have a lower reactivity with Pd than SiC. Post-test transmission electron microscope, TEM, examination showed that a $TiPd_x$ phase formed at the TiC/Pd interface, with $x \leq 3$. For the TiN/Pd interface, only FCC Pd(Ti) solid solutions were observed.

In general, the $M_{n+1}X_n$ layers in the MAX phases are quite stable in terms of reaction. In many cases, the preferred reaction path is one where the A-group element reacts selectively with the environment and out-diffuses from the MAX phase into the surrounding medium, leaving behind the binary carbides. Zhang et al.[41] studied the reactivity of $Ti_3AlC_2$ with Cu powders in the 800-1050°C temperature range. At lower temperatures there was little to no reaction. At 950°C and higher temperatures the reaction of $Ti_3AlC_2 + Cu = 3TiC_{0.67} + Al(Cu)$ was observed. It was proposed that Al deintercalated from the basal planes by diffusing out of MAX and forming solid solution with Cu - leaving behind nonstoichiometric $TiC_{0.67}$. In another study on the same $Ti_3AlC_2$-Cu system, Nechiche et al.[42,43] demonstrated that incorporation of Cu into the A site is accompanied by lattice distortion, which leads to symmetry reduction from a hexagonal to a monoclinic structure. It was also established that the variation of the lattice parameter of the solid solution compounds follows Vegard's law.

Furthermore, the A-group element may form intermetallic as was observed by Yin et al.[44] in their study of the reaction between $Ti_3SiC_2$ and Ni in the 800-1100°C temperature range. The diffusion path was determined to be $Ni/Ni_{31}Si_{12} + Ni_{16}Ti_6Si_7 + TiC_X/Ti_3SiC_2 + Ti_2Ni + TiC_X/Ti_3SiC_2$. Alternatively, a reaction by simple dissolution of the A-group element into the reacting medium may occur - as in the case of Ti and Nb in $Ti_2AlC$-$Nb_2AlC$ couples, studied by Ganguly et al.[45] In case that no intermetallic forms and the A-group element has no solubility in the reacting medium (e.g., Mg[46], Pb,[23] Pb–Bi[24]), then no reaction will be observed. In a recent study by Fashandi et al.,[47,48] new noble metal base MAX phases were discovered ($Ti_2AuC_2$, $Ti_3Au_2C_2$ and $Ti_3IrC_2$). For the formation of $Ti_2AuC_2$ the synthesis was done by a substitutional solid-state reaction of Au into $Ti_3SiC_2$ with simultaneous out-diffusion of Si from the A-layer. In case of $Ti_3IrC_2$ the reaction was done by substitution of Ir for Au in $Ti_3AuC_2$. These newly phases remained stable even after 1,000 h of annealing at 600°C in air. The group of the same authors previously reported on the synthesis of other MAX phases (($Cr_{0.5}Mn_{0.5})_2AuC$, $Mo_2AuC$, $Mo_2(Au_{1-x}Ga_x)_2C$ and $Mo_2(Ga,Fe,Au)C$), based on the same thermally induced substitutional reaction in the A-layer.[49-51]

The purpose of this work was to determine the reactivity of Pd with some common MAX phases. The ultimate question being: could the MAX phases in principle extend the lifetime of TRISO fuel pellets, by being used as an additional layer that maintains good thermal conductivity, or as a substitute to the current used SiC layer. In the present study, we focus on the interaction of $Ti_2AlC$, $Ti_3AlC_2$, $Ti_3SiC_2$ and $Cr_2AlC$ polycrystalline samples with Pd at 900°C.

**II. Experimental Details.**

The Ti$_2$AlC samples were sintered from Ti$_2$AlC powders (-325 mesh, Kanthal, Hallstahammar, Sweden) by hot pressing at 1300°C for 4 h under uniaxial pressure of about 36 MPa. In all cases, the vacuum level in the hot press, HP, was < 1 Pa and the die material was graphite. The Ti$_3$AlC$_2$ samples were prepared by hot pressing the same Ti$_2$AlC used above mixed with TiC powders (2 µm powder, 99.5%, Alfa Aesar, Ward Hill, MA) in a 1:1 ratio to get the right 3:1:2 stoichiometry of the Ti$_3$AlC$_2$ phase. The mixing was done by ball milling for 24 h and heating in the HP to 1400°C at a rate of 500 °C/h. The sample was held for 4 h under a load corresponding to a pressure of about 36 MPa, before furnace cooling. The Ti$_3$SiC$_2$ samples were prepared by hot pressing of Ti$_3$SiC$_2$ powders (-325 mesh, Kanthal, Hallstahammar, Sweden) for 4 h at 1500°C under vacuum < 1 Pa and a load corresponding to a pressure of about 36 MPa. Finally, the Cr$_2$AlC samples were prepared by ball milling a 2:1:1 stoichiometric mixture of Cr (-325 mesh, 99%, Alfa Aesar, Ward Hill, MA), Al (-325 mesh, 99.5%, Alfa Aesar, Ward Hill, MA), and graphite (-325 mesh, 99%, Alfa Aesar, Ward Hill, MA) powders for 24 h. Then, the mixture was loaded into a graphite mold and HPed for 4 h at 1400°C, under a load corresponding to a pressure of ~ 36 MPa.

The HPed samples were cut into a 0.3x1.5x1.5 cm$^3$ rectangles. Pd foil (1 mm thick, 99.99%, Alfa Aesar, Ward Hill, MA) was cut using electric discharge machining (EDM) into 0.5x0.5x0.1 cm$^3$ parallelepipeds. The Pd coupons were placed on top of the MAX phases in the same HP used to fabricate the samples and heated to 900 °C for 2 h, 4 h, and 10 h, under a load corresponding to a pressure of ~20 MPa and vacuum level lower than 1 Pa. After cooling, the samples were cross-sectioned, mounted, ground, polished and etched for 2-3 s by a 1:1:1 part solution of hydrofluoric acid (48%, Sigma Aldrich, St. Louis, MO), nitric acid (68%, Alfa Aesar, Ward Hill, MA) and water. The microstructure was examined by an optical microscope (OM) and a scanning electron microscope (SEM, FEI Zeiss Supra 50VP, Jena, Germany) equipped with an energy-dispersive X-ray spectroscope (EDS, Oxford INCA X-Sight 7573, Abingdon, England). In all cases, except the $Cr_2AlC$ composition, etching resulted in brilliant colours under the OM.

X-ray diffraction (XRD) patterns of both sides of the fractured interface after thermal treatment of each diffusion couple were obtained using a diffractometer (Rigaku SmartLab, Tokyo, Japan) with a 2 or 5 mm slit, depending on sample size. The scans were conducted over the range of 5 to 70° 2θ, with a step size of 0.02° and a dwell time of 0.5 s per step.

## III. Results and Discussion

When the XRD patterns of the interfaces of the various diffusion couples are plotted (not shown) the only definitive result was obtained on the $Cr_2AlC$ side of its diffusion couple, clearly identifying PdAl as a reaction phase (Fig. 1). Several attempts were made to obtain similar information for the other diffusion couples, but no useful information was gleaned, most probably because the fractured surfaces were quite rough or/and the volume fraction of the reaction product was too small. Furthermore, if the fracture plane does not cleave the diffusion couple exactly where the reaction layer is located, the likelihood of obtaining an XRD signal is not high. Nevertheless, the definitive evidence for PdAl being a reaction phase in the $Cr_2AlC$/Pd system is important since (as discussed below) the EDS results suggest it to be the reaction product in the $Ti_2AlC$/Pd and $Ti_3AlC_2$/Pd systems, as well.

In the following, the microstructures of each diffusion couple are presented separately.

### **$Cr_2AlC$/Pd system**

Backscatter electron, BSE, SEM micrographs of the $Cr_2AlC$/Pd interface after heating to 900°C for 2, 4, and 10 h shown in Fig 2a, b, and c, respectively, show the growth of the diffusion layer from 20 µm, after 2 h, to 45 µm, after 10 h.

A BSE SEM micrograph of the interface after heating to 900°C for 10 h is shown in Fig. 3a. The corresponding elemental maps of Al, Pd, and Cr are shown in Figs. 3b, c, and d, respectively. From these micrographs it is obvious that both Pd and Al migrate simultaneously, while Cr diffusion is quite limited. The presence of small dark regions in the reaction layer in Fig. 3a, that appear to be similar in shape and morphology to those in the underlying $Cr_2AlC$ suggest that the location of the original interface is given by the red line in Fig. 3.

**Ti$_2$AlC/Pd system**

Fig. 4a is a BSE SEM micrograph of the Ti$_2$AlC/Pd interface, after heating to 900°C for 2 h. Fig. 4b is a higher magnification OM micrograph of the same region. From this figure it is clear that a reaction layer forms on the Pd side. BSE SEM micrographs of the Ti$_2$AlC/Pd interface, after heating to 900°C for 4 and 10 h are shown in Figs. 4c and 4d, respectively. These micrographs clearly show that Pd penetrated to a depth of ~ 220 µm into the Ti$_2$AlC. Interestingly, the penetration depth is *independent* of time, or at most, a weak function of it. The reason for this state of affairs is unclear at this time. One possibility is that the penetration/reaction is surface reaction rate controlled and thus time independent. According to the binary Pd-Al phase diagram the eutectic temperature is about 1055°C.[52] If one assumes that Ti and C dissolution into Pd leads to a further decrease in the eutectic temperature, then it is possible that a liquid phase existed at 900 °C. This microstructure shown - unique to the Al-containing phases - is consistent with the presence of a liquid penetrating along the grain boundaries. This microstructure is not observed for the other two couples. In addition to Pd penetration, two diffusion layers are discernible: a 35 µm thick on the Ti$_2$AlC side of the interface and a 20 µm thick on the Pd side (Figs. 4).

A BSE SEM micrograph of the interface and elemental maps of Al, Pd and Ti across that interface are shown in Fig. 5. In this case there is no ambiguity as to the location of the original interface since the outline of the Ti$_2$AlC grains are clearly visible to the left of the thin red line in Fig. 4b. The elemental maps revealed that the Pd atoms diffuse into the MAX phase and Al atoms diffuse into Pd. The Ti atoms do not appear to diffuse.

Based on EDS point analysis (not shown) of various the regions shown in Fig. 5, the following phases, in addition to the original, have been identified. On the right side of the red line, a Pd-Al intermetallic phase, with a Pd:Al ratio of 4:3. On the left side of the red line, in the vicinity of the interface, the composition of the continuous newly formed layer corresponds to the TiPd$_3$ phase. The bright areas in the infiltrated region are Pd:Al with an atomic ratio close to 1:1.

**Ti$_3$AlC$_2$/Pd system**

Fig. 6a is an OM micrograph of the Ti$_3$AlC$_2$/Pd interface, after heating to 900°C for 2 h. Fig. 6b is a higher magnification micrograph of same region. From these figures it is clear that a reaction layer forms on the Pd side. BSE SEM micrographs of the Ti$_3$AlC$_2$/Pd interface, after heating to 900°C for 4 h and 10 h are shown in Fig. 6c and d, respectively. These micrographs clearly show that the composition and morphology of the regions close to the interface is similar to those found in the Ti$_2$AlC/Pd system. BSE SEM micrographs of the Ti$_3$AlC$_2$/Pd interface after heating to 900°C for 2 and 10 h, presented in Fig. 6c and d, respectively, show that the diffusion layer grows from 70 µm after 2 h to 105 µm after 10 h, respectively.

A BSE SEM micrograph of the interface (Fig. 7a) along with elemental maps of Al, Pd, and Ti are presented in Fig. 7b, 7c, and 7d, respectively. The interface shows similar features to the Ti$_2$AlC/Pd interface case at the same conditions. Here again there is no ambiguity as to the location of the original interface since the outline of the Ti$_3$AlC$_2$ grains are clearly visible to the left of the thin red line in Fig. 6b. The diffusion layer clearly shows the migration of Pd, and the depletion of Al on the Ti$_3$AlC$_2$ side (Fig. 7b). EDS point analysis suggest on a formation of a palladium aluminide phase on the Al rich region viz. on the right side of the red line (Fig. 7b). It is thus reasonable to conclude that in this diffusion couple the Pd and Al are inter-diffusing, while the Ti appears to be immobile. It is important to point here that Pd penetration deep into the Ti$_3$AlC$_2$ substrate is also evidenced by the Pd map shown in Fig. 7c. The morphology of this phase is not as clear as in the Ti$_2$AlC case, but it is reasonable to assume that the Pd in this case also penetrated along the grain boundaries through possibly a liquid phase. That the thickness of this layer - at ≈ 200 µm – is comparable to that in the Ti$_2$AlC case, is thus probably not a coincidence.

By assuming that same reaction takes place between Pd and other Al-based MAX phase, the next general reaction can be suggested:

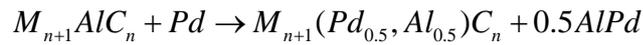

$$M_{n+1}AlC_n + Pd \rightarrow M_{n+1}(Pd_{0.5}, Al_{0.5})C_n + 0.5AlPd$$

### Ti$_3$SiC$_2$/Pd system

Figure 8a shows a cross-sectional polished and etched OM micrograph of the Ti$_3$SiC$_2$/Pd interface after heating to 900°C for 2 h. Figure 8b shows the area depicted by the rectangle in Fig. 8a at higher magnification. BSE SEM micrographs of the interfaces obtained after heating for 4 and 10 h, are shown in Figs. 8c and 8d, respectively. Note that after 10 h (Fig. 8d), the diffusion couple separated near the original interface, allowing epoxy, so labelled, to penetrate the interface during mounting of the sample.

A BSE SEM micrograph of this interface is shown in Fig. 9a. Elemental maps of Si, Pd and Ti across the interface are shown Figs. 9b, c, and d, respectively. Based on these maps - where the original interface is depicted by a vertical thin red line - it is obvious that the Pd atoms diffuse into the MAX phase, the Si atoms diffuse into the Pd (see below), whereas the Ti atoms do not appear to diffuse. Unlike the previous cases of Ti$_2$AlC and Ti$_3$AlC$_2$, however, there are no clear signs of a Pd based melt infiltrating into the MAX sample. This is surprising since the eutectic temperature in the Pd-Si system is lower than 900°C.[53] The thickness of the layer on the left side of the original interface (Fig. 8) increased from 200 µm after 2 h to 410 µm after 10 h of annealing.

Based on Figs. 8 and 9 combined with EDS point analysis, the following regions, in addition to the original pure Pd and Ti$_3$SiC$_2$, were identified:

a) A ≈ 10-20 µm thin layer, near the original interface, best seen, and labelled i, in Figs. 8a and 8b. According to EDS analysis the atomic Pd:Si ratio was 0.68±0.1:0.32±0.02, or ≈ $Pd_{2.14}Si$. b) A thick layer – whose thickness is ≈ 200, 240 and 410 µm after 2, 4 and 10 h of annealing, respectively - is clearly observed on the left side of the original interface, and labelled ii, in Figs. 8a and 8b. According to EDS analysis of this region, the Ti:Si:Pd atomic ratios were 0.67±0.1:0.15±0.05:0.18±0.1, or ≈ $Ti_2Si_{0.45}Pd_{0.54}$. The whitish grains in the etched OM micrographs are TiC particles.

To shed more light on the nature of this region, a lamella of that layer was focus ion beam machined for TEM observations. Typical TEM results of that layer are shown in Fig. 10a, from which it is clear that the hexagonal symmetry (Fig. 10b and 10c) and layered MAX phase structure are maintained. EDX of 5 different regions resulted in the following Ti:Si:Pd atomic fractions: 0.49±0.07:0.23±0.06:0.26±0.06 or ≈ $Ti_2Si_{0.5}Pd_{0.56}$. It follows that region ii in Fig. 8a is most probably a MAX phase solid solution with Pd. If one assumes a reduced Si content to account for the Si that diffuses into the Pd, this ratio translates to a solid solution of Pd in $Ti_3SiC_2$, with an approximate chemistry of $(Ti_{1.95},Pd_{1.05})(Si_{0.9},Pd_{0.1})C_2$. This result is intriguing because recent work has shown that when the MAX phases are placed in the vicinity of Au and Cu the latter substitute for the A element. That was also the case in the present study for the $Ti_2AlC$/Pd, $Ti_3AlC_2$/Pd and $Cr_2AlC$/Pd couples. However, in regard the $Ti_3SiC_2$/Pd system, a more complex substitution reaction take place, where about one third of the Ti atoms in the original M layer are replaced by Pd. The out-diffused Ti atoms migrate to the pre-excited $TiC_z$ particles (appear as white particles on the OM micrograph in Figs. 8a and 8b) in the MAX phase, leading to their growth and stichometry change. This is consistent with the XRD analysis (not shown), where the lattice parameter of the pre-existing $TiC_z$ particles shifts after the HP treatment from 4.327 to 4.317 Å. That change in the lattice constant correspond to a change in stoichiometry of $z$ from 0.9 to 0.6 respectively.[54] From the present results, we suggest that the following reaction takes place:

$$Ti_3SiC_2 + TiC + Pd_{1+3y} \rightarrow (Ti_2, Pd)(Si_{1-y}, Pd_y)C_2 + TiC_{0.5} + yPd_2Si.$$

When the thickness, $x$, of the newly formed diffusion layer on the MAX side (Table 1) that is time dependent (e.g. shown by white line in Figs. 8c and 8d) is plotted vs. $t^{0.5}$ (Fig. 11), a straight line passing through the origin is obtained, suggesting diffusion limited kinetics. Least squares fitting of the lines yield correlation coefficients higher than 0.98 in all cases. By assuming that the process is diffusion controlled, viz. $x^2 = (Dt)^{0.5}$, where $D$ is a diffusion coefficient of Pd in MAX and $t$ is time, a $D$ of $2.15 \times 10^{-12}$, $6.16 \times 10^{-13}$, $2.4 \times 10^{-13}$ and $1.61 \times 10^{-13}$ m$^2$/s were obtained for the Ti$_3$SiC$_2$/Pd, Ti$_3$AlC$_2$/Pd, Cr$_2$AlC/Pd and Ti$_2$AlC/Pd systems respectively. Thus, the most resistant MAX phase to reaction with Pd, is Ti$_2$AlC and the least is Ti$_3$SiC$_2$, with Cr$_2$AlC and Ti$_3$AlC$_2$, in between.

**Summary and Conclusions:**

Herein we report, for the first time, on the interaction between Pd with the MAX phases, Cr$_2$AlC, Ti$_2$AlC, Ti$_3$AlC$_2$ and Ti$_3$SiC$_2$. Table 1 summarizes the results after the diffusion couples were held at 900°C for 10 h. In all cases, Pd diffused into, and the A-element diffused out of, the MAX phase. In the case of Cr$_2$AlC, Al diffusion into Pd resulted in the formation of the intermetallic AlPd. Not too surprisingly, Ti$_2$AlC and Ti$_3$AlC$_2$ behaved similar to each other and shared similarities with the reaction between TiC and Pd where an infiltration of Pd into the TiC occured.[39] Migration of Al into the Pd was also detected, suggesting the formation of a TiPd$_3$ phase. And while there is evidence for rapid liquid penetration - ≈ 200 µm - into Ti$_2$AlC and Ti$_3$AlC$_2$, such evidence was not found in Cr$_2$AlC or Ti$_3$SiC$_2$. Interestingly and for reasons that are unclear, the thickness of the liquid penetration was not a function of time.

$Ti_3SiC_2$ behaved more like SiC,[35-38] than TiC,[39,40] showing a quite rapid diffusion of Pd into the MAX phase. Similar to Al in $Ti_2AlC$ and $Ti_3AlC_2$, and Si in $Ti_3SiC_2$ diffuses into Pd, leading to the formation of a $Pd_2Si$ phase. The interaction between $Ti_3SiC_2$ and Pd was also found to form a MAX phase solid solution, with an approximate chemistry of $(Ti_{1.95},Pd_{1.05})(Si_{0.9},Pd_{0.1})C_2$.

After heating to 900°C for 10 h, the time dependable diffusion layer thickness on the MAX phase side in $Ti_2AlC$/Pd, $Ti_3AlC_2$/Pd, $Ti_3SiC_2$/Pd, and $Cr_2AlC$/Pd was found to be 35, 105, 410, and 45 µm, respectively. The values of the diffusion coefficients of Pd in various MAX phases were estimated. It was established that the best resistance to Pd attack follows, from best to worst, $Ti_2AlC$, $Cr_2AlC$, $Ti_3AlC_2$, and $Ti_3SiC_2$. However, given the ≈ 220 µm penetration of what appear to be a liquid phase into the grain boundaries in the case of $Ti_2AlC$ and $Ti_3AlC_2$, the most stable phase in terms of Pd penetration is $Cr_2AlC$. This conclusion notwithstanding, given the relatively low times and temperatures used herein, it is clear that these solids would make poor diffusion barriers for Pd at temperatures in the vicinity of 900°C.

**Acknowledgements:**

This research has received funding from the Euratom research and training programme 2014–2018 under grant agreement No. 740415 (H2020 IL TROVATORE).

**Tables**

**Table 1.** Summary of newly formed phases after heat treatment at 900°C for 10 h.

| Diffusion couple | MAX side | | Pd side | | Liquid penetration into MAX | |
|---|---|---|---|---|---|---|
| $Cr_2AlC$/Pd | $Cr_2(Al,Pd)C$ | 45 μm | AlPd | <10 μm | no evidence | |
| $Ti_2AlC$/Pd | $Ti_2(Al,Pd)C$ | 35 μm | AlPd | 20 μm | TiAl | ~220 μm (time independable) |
| $Ti_3AlC_2$/Pd | $Ti_3(Al,Pd)C_2$ | 105 μm | AlPd | 30 μm | TiAl | ~220 μm (time independable) |
| $Ti_3SiC_2$/Pd | $(Ti_{1.95},Pd_{1.05})(Si_{0.9},Pd_{0.1})C_2$ | 410 μm | $Pd_{2.14}Si$ | 20 μm | no evidence | |

**Figures**

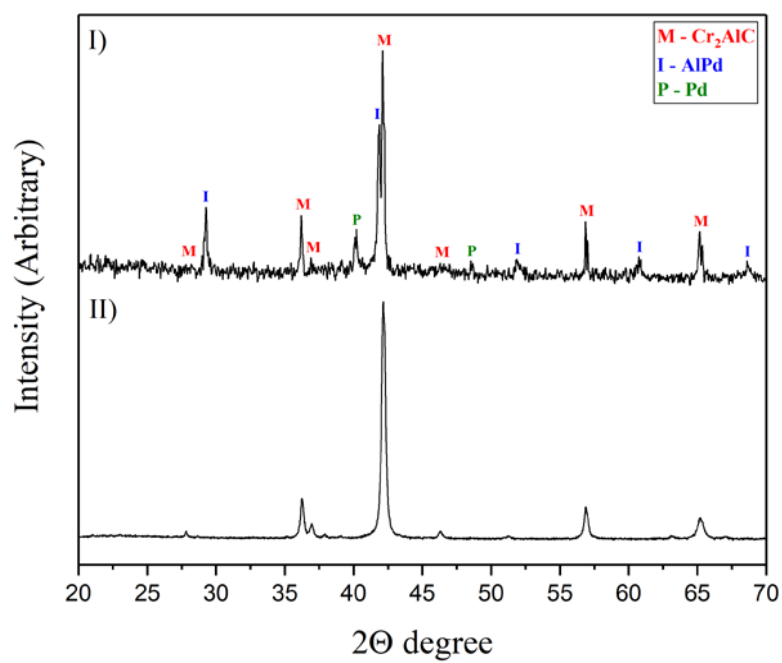

**Fig. 1.** XRD patterns of the interface between I), Cr$_2$AlC and Pd after heating at 900 °C for 10 h and, II), control

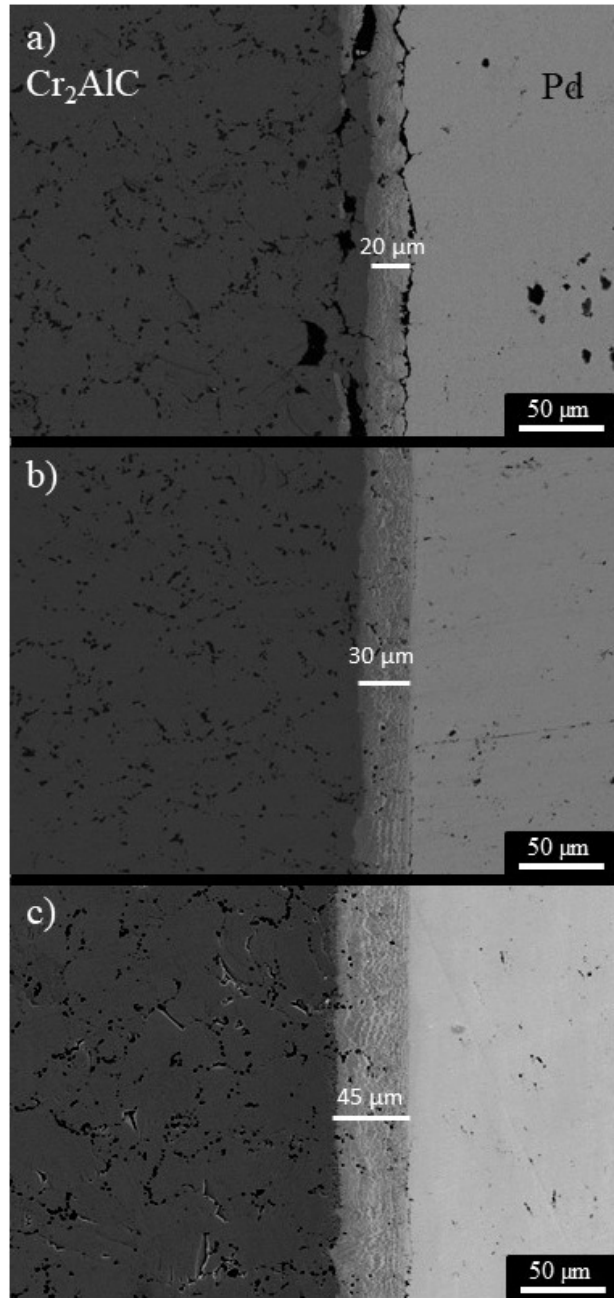

**Fig. 2.** Backscatter electron SEM micrographs of the Cr$_2$AlC/Pd interface after heating to 900°C for, a) 2, b) 4, and, c) 10 h. Black particles on the MAX side correspond to Al$_2$O$_3$.

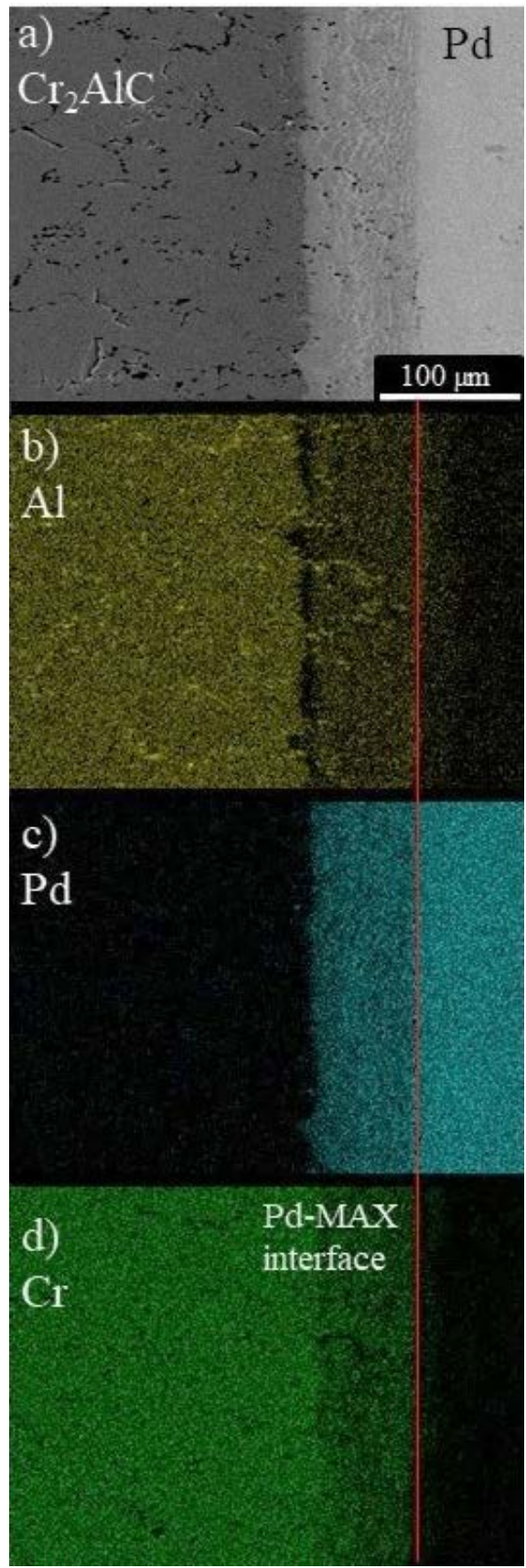

**Fig. 3.** a) Backscatter electron SEM micrographs of the $Cr_2AlC$/Pd interface after heating to 900°C 10 h, with elemental mappings of, b) Al, c) Pd, and, d) Cr. Thin red line is approximate location of original interface.

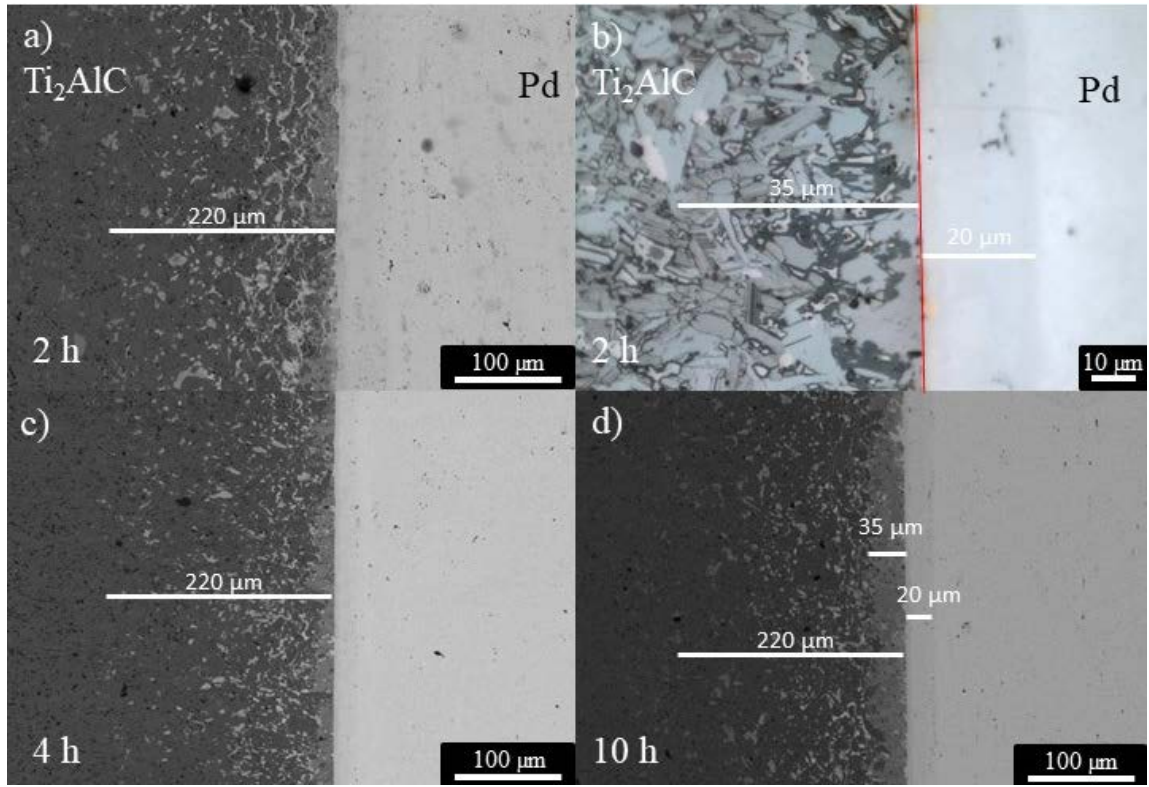

**Fig. 4.** Ti$_2$AlC/Pd interface after heating to 900°C, a) BSE SEM micrograph after 2 h, b) OM micrograph of the etched surface after 2h, c) BSE SEM micrograph after 4 h, and, c) BSE SEM micrograph after 10 h. Red line in b denotes original interface.

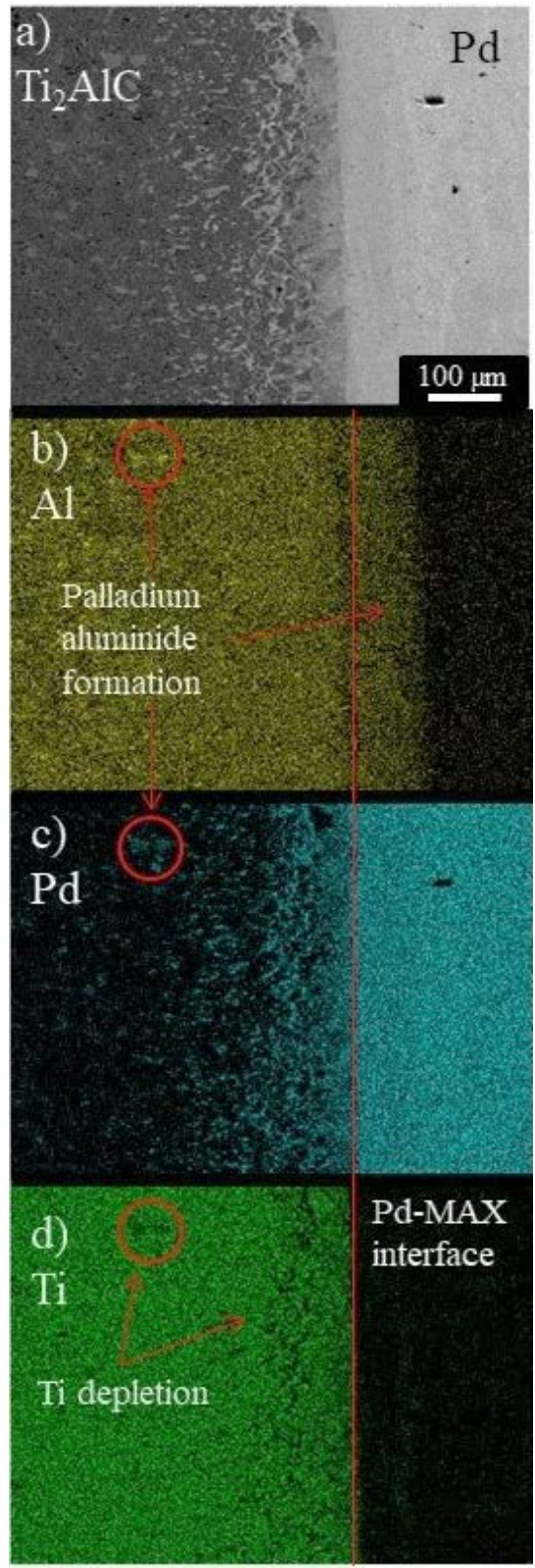

**Fig. 5.** a) Backscatter electron SEM micrographs of the Ti$_2$AlC/Pd interface after heating to 900°C 10 h, with elemental mappings of, b) Al, c) Pd, and, d) Ti. Red line denotes location of original interface.

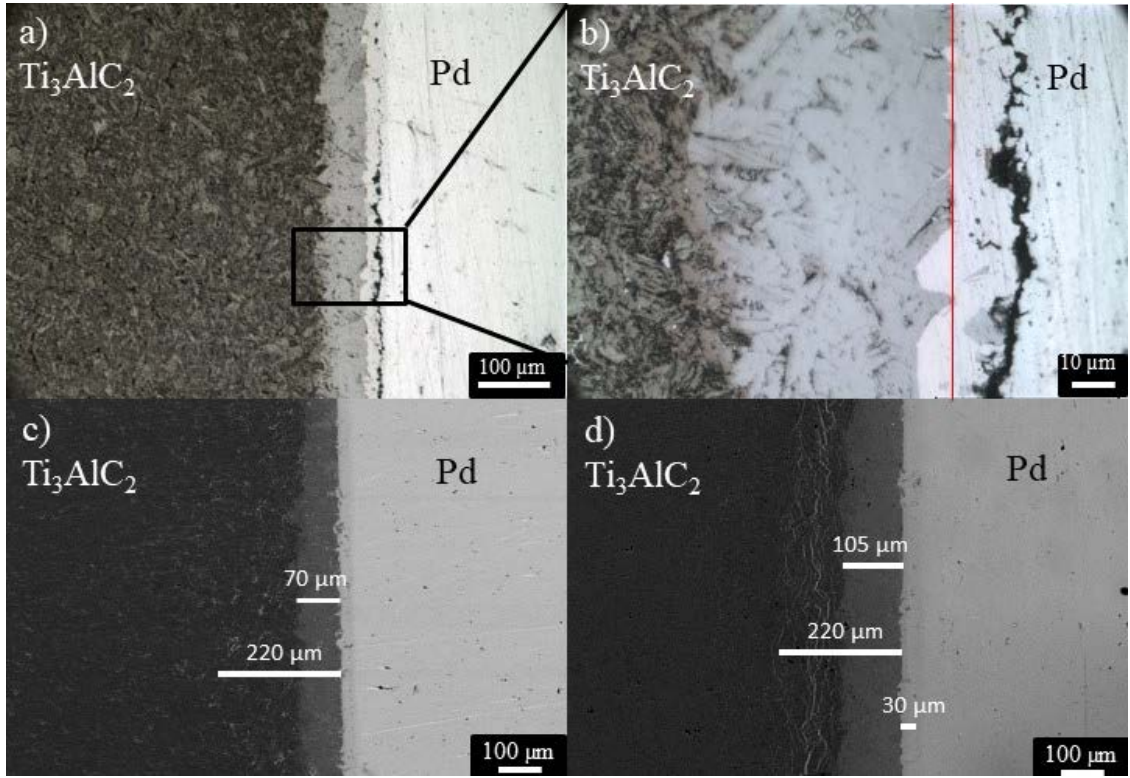

**Fig. 6.** Ti$_3$AlC$_2$/Pd interface after heating to 900°C, a) OM micrograph of the etched surface after 2 h, b) OM micrograph at higher magnification, c) BSE SEM micrograph after 2 h and, d) after 10 h. Red line in b shows location of original interface. Note fracture plane to the right of the line.

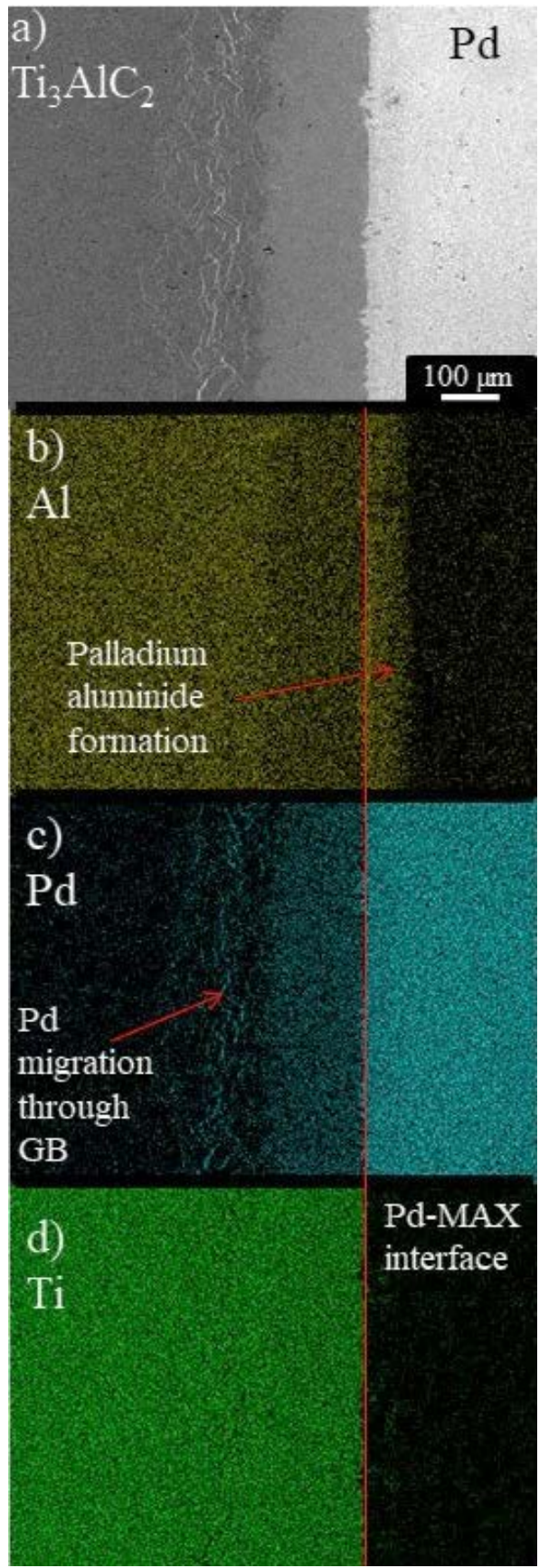

**Fig. 7.** a) Backscatter electron SEM micrographs of the Ti$_3$AlC$_2$/Pd interface after heating to 900°C 10 h, with elemental mappings of, b) Al, c) Pd, and, d) Ti. Red line denotes location of original interface.

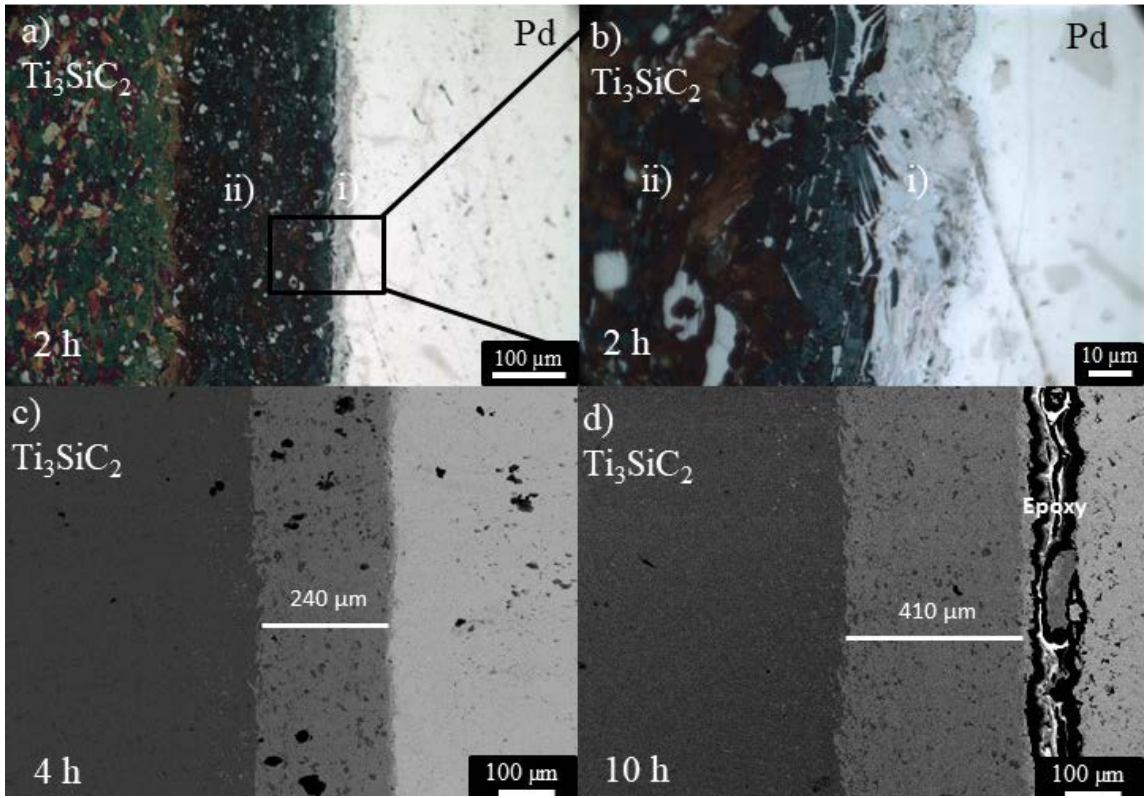

**Fig. 8.** Ti$_3$SiC$_2$/Pd interface after heating to 900°C, a) OM micrograph of the etched surface after 2 h, and b) at higher magnification, c) BSE SEM micrograph after 4 h, and, d) BSE SEM micrograph after 10 h.

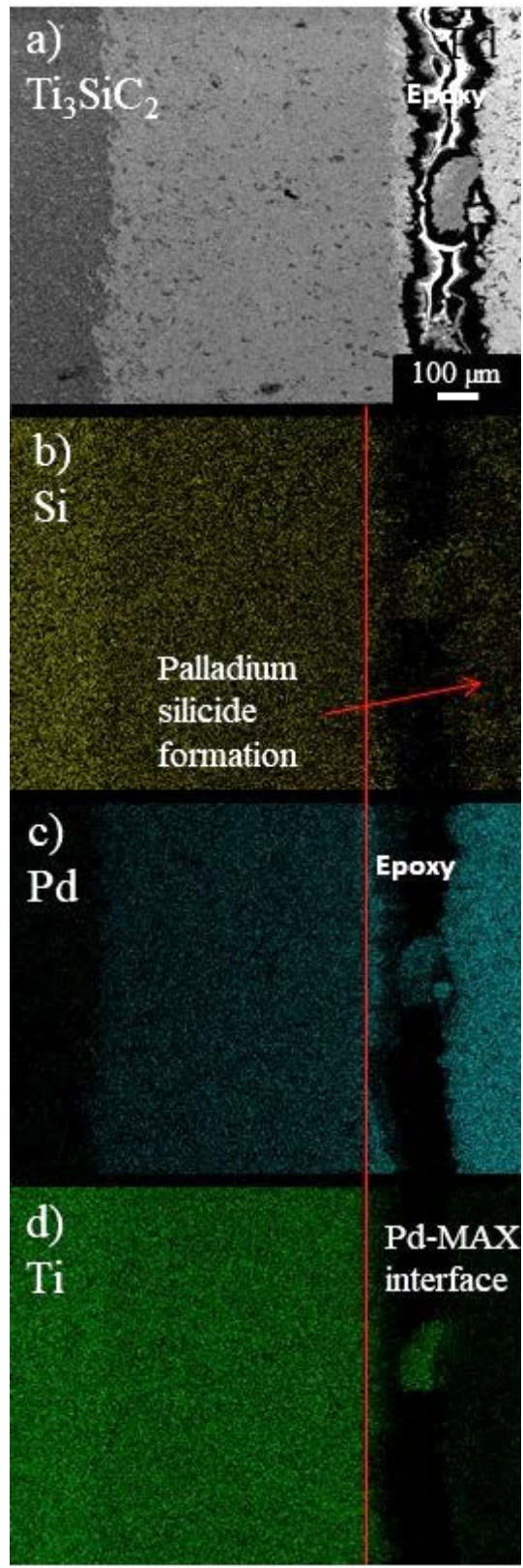

**Fig. 9**. a) Backscatter electron SEM micrographs of the Ti$_3$SiC$_2$/Pd interface after heating to 900°C 10 h, with elemental mappings of, b) Si, c) Pd, and, d) Ti. Red line denotes location of original interface.

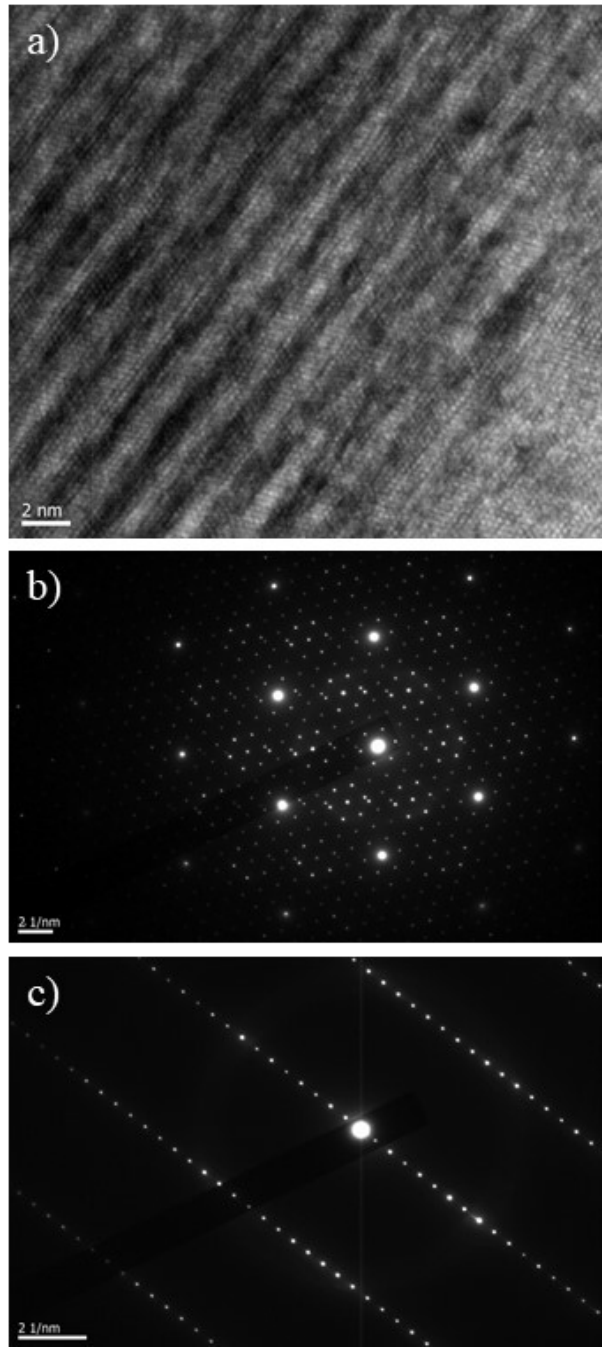

**Fig. 10.** a) HRTEM of region ii in Fig. 8a from the Ti$_3$SiC$_2$/Pd couple heated to 900°C for 2 h. Diffraction patterns along, b) [11$\bar{2}$1] and, c) [11$\bar{2}$0] directions.

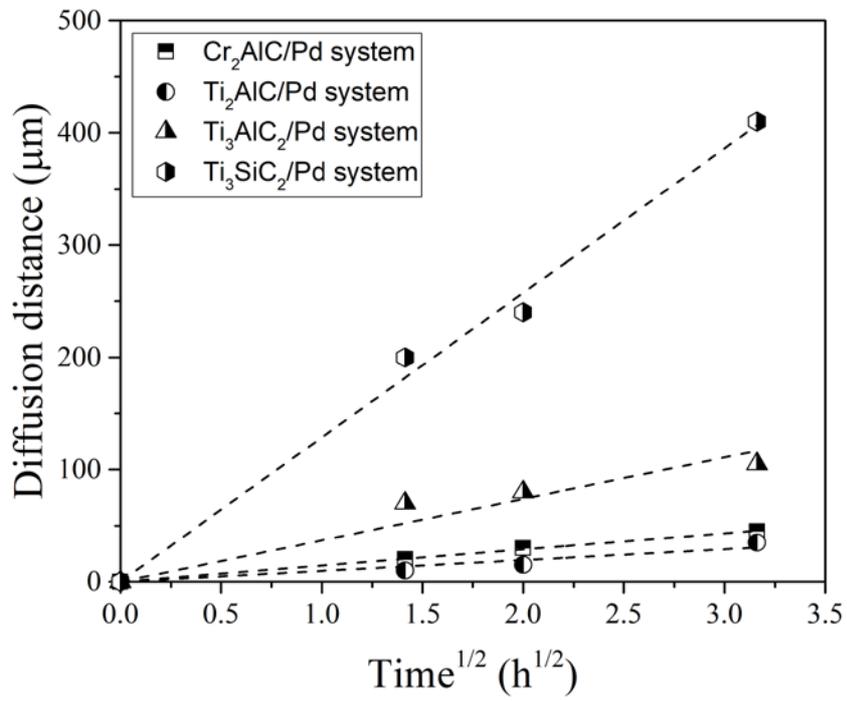

**Fig. 11.** Thickness of interfacial layers growing into the MAX phase for the various MAX/Pd couples held at 900 °C as a function of √t.